\documentclass{llncs}

\usepackage{enumerate}

\usepackage{graphicx,amsmath, amsfonts}
\usepackage[usenames]{color}
\usepackage[english]{babel}
\usepackage[utf8]{inputenc}
\usepackage[T1]{fontenc}
\usepackage{fancyhdr}
\usepackage{color}
\usepackage{amssymb}

\newlength\shadedboxwidth
\long\def\shaded#1{\begin{trivlist}\item[]%
\setlength\fboxsep{2ex}
\setlength\fboxrule{.4pt}
\def\bgcolor{.99}
\setlength\shadedboxwidth\linewidth
\addtolength\shadedboxwidth{-2\fboxsep}%
\addtolength\shadedboxwidth{-2\fboxrule}%
\fcolorbox[gray]{0}{\bgcolor}%
{\parbox{\shadedboxwidth}{#1}}\end{trivlist}}

\newtheorem{exercisee}[lemma]{Exercise}
\newtheorem{theoremm}[lemma]{Theorem}

\newcommand{\eop}{\hfill $\square$\\}
 
\author{Tomasz Gogacz, Jerzy Marcinkowski}
\institute{Institute of Computer Science,\\ University of Wrocław}
\title{All--instances termination of chase is undecidable}

\begin{document}
\maketitle
\pagestyle{headings}
 
\begin{abstract} We show that all--instances termination of  chase is undecidable. 
More precisely, there is no algorithm deciding, for a given set $\cal T$ consisting
of Tuple Generating Dependencies (a.k.a. Datalog$^\exists$ program), whether the $\cal T$-chase on $D$ will terminate for every 
finite database instance $D$.
Our method  applies to Oblivious Chase, Semi-Oblivious Chase and -- after a slight modification -- also for Standard Chase.
This means that we give a (negative) solution to the all--instances termination problem for all version of chase that are usually considered. 

The arity we need for our undecidability
 proof is three. We also show that the problem is EXPSPACE-hard for binary signatures, but decidability for this case is left open.

Both the proofs -- for ternary and binary signatures -- are easy.  Once you know them.

\end{abstract}

\section{Introduction}

The chase procedure was defined in late 1970s and has been considered one of the most fundamental database theory algorithms since then.
It has been applied to a wide spectrum of problems, for example for
checking containment of queries under constraints [ASU79] or for  testing  implication between sets of database dependencies ([MMS79], [BV84]). 
A new wave of interest in this notion
began when the theory of data integration was founded ([FKPP05]), where chase is used to
compute solutions to data exchange problems. This interest was 
further strengthened recently by the Datalog$^\pm$ program [CGL09], [CGL12].\smallskip

The {\bf basic idea of a $\cal T$-chase} is as follows. We consider a  set $\cal T$ of Tuple Generating 
Dependencies\footnote{Such sets are also known as Datalog$^\exists$ programs, and we will use the word ``program'' in this sense. While chase is sometimes also defined for other types of dependencies, we only consider Tuple Generating Dependencies
 in this paper.},  which means rules (constraints) of the form:

$$ \Phi(\bar x,\bar y)\Rightarrow \exists  \bar z \;\Psi(\bar x,\bar z)$$

where $\Phi$ and $\Psi$ are conjunctive queries\footnote{$\Phi$ and $\Psi$ are positive, without equality. Our negative results hold for single head TGDs, which means that $\Psi$ is a single atom.}, and where $\bar x$, $\bar y$ and $\bar z$ are tuples of variables. Then, for 
a database instance $D$ we try -- step by step -- to extend $D$, by adding new elements and atoms,
 so that the new database satisfies the constraints from $\cal T$: {\em whenever} there
 are some elements $\bar a$, $\bar b$ in the current structure,  such that  $\Phi(\bar a, \bar b)$ is true, a tuple $\bar c$ of new elements is created and new relational atoms added,
to make $\Psi(\bar a,\bar c)$ also true. Notice that the tuple $\bar z$ can be empty. In such case the TGD under consideration 
degenerates to a plain Datalog rule. 

As it turns out, there are several possible semantics of the {\em whenever} above, leading to several versions of the chase procedure. The
Standard Chase is a lazy version  -- it only adds new elements if  $\Phi(\bar a,\bar b)$  is true in the current structure, but
$\exists \bar z \; \Psi(\bar a,\bar z)$ is  (at this point of execution) false. Oblivious and Semi-Oblivious Chase ([M09]) are eager versions. 
Oblivious Chase always adds one tuple $\bar c$ for each tuple 
$\bar a, \bar b$ such that  $\Phi(\bar a, \bar b)$ is true. Semi-Oblivious Chase always adds one tuple $\bar c$ for each tuple $\bar a $ such that  
$\exists \bar y\; \Phi(\bar a, \bar y)$ is true.

It is not hard to notice that the order of execution does not matter for  Oblivious and Semi-Oblivious Chase. Whatever order the candidate tuples are picked in, 
we will 
eventually get the same structure\footnote{If chase does not terminate the claim is true provided the order is fair -- each tuple will be eventually picked.}. 
But Standard Chase is non-deterministic -- different orders in which tuples are picked can eventually lead to different structures. 

One more version of the procedure is Core Chase (see [DNR08]). It is again a lazy version, but a parallel one: all the rules
applicable at some point are triggered at the same time. In this way the non-determinism of Standard Chase is got rid of. For reasons that we
will not discuss here Core Chase is slightly more complicated than that (and not really practical -- the cost of each step is DP-complete).\smallskip

As we said before, the chase procedure  is almost ubiquitous in database theory. This phenomenon is discussed in [DNR08]:  {\em ``the
applicability of the same tool to (..) seemingly different problems
is not accidental, and it is due to a deeper, tool--independent reason:
to solve these problems, it suffices to exhibit a representative} (database) {\em instance $U$ with two key properties, and the chase is an algorithm for
finding such an instance.''} The two key properties of the instance $U$, being the result of $\cal T$-chase on an a database instance $D$, 
for given set $\cal T$ of tuple generating dependencies 
and for given database instance $D$ are that:\smallskip

\noindent
- \hspace{-0.8mm}$U$ is a model of $\cal T$ and $D$;\\
- \hspace{-0.8mm}$U$ is universal - there is  a homomorphism from $U$ into every model of $D$ and $\cal T$.\smallskip

 But $U$, or {\bf Chase$(D,{\cal T})$}, as we prefer to call the structure resulting from running a $\cal T$-chase on $D$, is in many cases only useful when it is finite, which only happens  if (and only if) the chase procedure terminates. One of the applications where 
finiteness of Chase$(D,{\cal T})$ is a key issue is  considered in [FKPP05], and a sufficient 
condition on $\cal T$, implying finiteness of Chase$(D,{\cal T})$ 
was studied in this paper, called  Weak Acyclicity. 
Weak Acyclicity is a property of $\cal T$ alone, so it implies termination regardless of  $D$. 
This reflects the fact that the typical context in which  database constraints are analyzed is the static analysis context -- we want to optimize $\cal T$ 
before knowing $D$.  So, in particular,  it is natural to want to be sure that  $\cal T$--chase on $D$  will terminate on $D$ before knowing the $D$ itself.
 Many other conditions  like that were studied. For example the Stratified-Witness property ([DT03]), 
which is historically earlier, and stronger (i.e. narrower), than Weak Acyclicity. Then it was the Rich Acyclicity criterion, introduced in [HS07], 
and proved in [GO11]
to imply termination of Oblivious
Chase for all instances $D$. 
A condition based on stratification of rules was introduced in [DNR08]. As it turned out to only guarantee termination of the Standard Chase, 
another class of sets of rules -- Corrected Stratified Class (CSC) was defined in [MSL09], with Oblivious Chase terminating for all instances $D$. Then, in [MSL09a], 
CSC was extended to  Inductively Restricted ($IR$) class, and further to a whole hierarchy of classes $T[k]$, where  $T[2]=IR$.

This list is by no means exhaustive -- see Adrian Onet's thesis [O12] for a 35-pages long survey chapter about sufficient conditions for chase termination. What is
however worth mentioning is that all the known conditions imply all-instances termination and thus none of them depends on $D$.

With so much effort spent on finding the sufficient conditions it is  natural to ask about decidability of the all--instances termination problem itself. 
But surprisingly, this fundamental problem has so far remained
open. Some work was done, but mostly on a related  
problem of chase termination  for given program $\cal T$ and also {\bf given} database instance $D$.
It was shown to be undecidable in [DNR08] for Core Chase and Standard Chase ($\spadesuit$). In [M09] it was noticed that the proof of $\spadesuit$
 works also for Semi-oblivious and Oblivious chase. 
The only previous results concerning decidability  of the all--instances chase termination problem  can be found in [G013], where 
the problem is shown to be undecidable for Core Chase ($\heartsuit 1$)  and the Standard$_\exists$ sub-version ($\heartsuit 2$), where we ask, for given $\cal T$,  whether for each database instance $D$
{\bf there exists} a terminating execution path of  $\cal T$-Standard Chase on $D$  (let us remind here that Standard Chase is a non-deterministic procedure). 
And this is again not really the most natural question as -- having some $\cal T$ on mind -- we want to be sure that whenever and however we run a $\cal T$-chase, 
it will always terminate\footnote{The termination problem for the Standard$_\exists$ version is shown to be $\Pi^0_2$ complete in [GO13]. But the result statement there is not correct: co-r.e. completeness is claimed.}.

Another result in [GO13] is undecidability of all--instances chase termination problem for sets of constraints where, apart from TGDs, a denial constraint 
is allowed, which is a conjunctive query $Q$ such that when $Q$ is proved somewhere in Chase$({\cal T}, D)$ then the chase procedure terminates and ``fails'' ($\clubsuit$).

One more result from [M09], which can be slightly confusing, is undecidability of what is there -- misleadingly -- called ``all-instances termination'' ($\diamondsuit$). The signature $\Sigma$ of the TGDs there 
is a disjoint union of two sub-signatures $\Sigma_1$ and $\Sigma_2$ but only instances where the relations in $\Sigma_2$ are initially empty are allowed. 
\vspace{-1mm}

\subsection{Our contribution}

\vspace{-1mm}
The main result of this paper is:
\vspace{-1mm}

\begin{theoremm}\label{glowne}
All-instance termination of Oblivious Chase is undecidable (and r.e.-hard) 
 for programs consisting of single-head TGDs over ternary signatures.
\end{theoremm}

Proof of Theorem \ref{glowne} is  presented in Section \ref{dowod}. It  can also be read, without any changes,
  as a proof of  undecidability of all-instances termination of  Semi-Oblivious Chase. In short Subsection \ref{standard} we modify the proof to
show that also all-instances--all--paths Standard Chase termination is  undecidable\footnote{This is for sake of completeness, as it was earlier
 shown in  
 [GO13]  that undecidability of all-instances termination of  Oblivious Chase
implies  undecidability of  all-instances--all--paths Standard Chase termination.}.

It is common knowledge that whatever can be said about TGDs over high arity signature usually remains true for binary signature, as long as
multi-head TGDs are allowed.  And also the other way round -- one who is prepared to pay the arity cost can usually translate everything into the language of single-head TGDs. This fails however  in the context of chase termination: one can easily modify our proof of Theorem \ref{glowne} to get undecidability of
all-instance chase termination for multi-head binary TGDs, but only for  Semi-Oblivious Chase, not for Oblivious. See Appendix D. for details. 
In Section \ref{bibarne}  we show:

\begin{theoremm}\label{glowne3}
All-instance termination of Oblivious Chase is EXPSPACE-hard for programs consisting of   single head TGDs over binary signatures. 
\end{theoremm}

\noindent
{\bf Upper bounds.} It follows easily from Lemma \ref{studnia} that the all-instances termination problem of Oblivious and Semi-Oblivious Chase is recursively enumerable,
and so Theorem \ref{glowne} provides  matching lower bounds. 
But Lemma \ref{studnia} is not true for all--instances--all--paths Standard Chase termination,
and thus the only upper bound known for this problem is the $\Pi^0_2$ level of the Arithmetical Hierarchy. Our conjecture is that the problem is 
in fact also r.e., but much more insight into the structure of Standard Chase is needed in order to prove this claim.

The lower bound given by Theorem \ref{glowne3} is  not matched by any upper bound, and we believe that the problem is undecidable. 
A similarity that  is maybe worth being mentioned here 
(see also next subsection) -- is that Datalog programs uniform boundedness is 
also known to be undecidable for ternary arities but decidability was left open for the binary case [M99].
\vspace{-2.5mm}

\section{Techniques} 

\vspace{-1mm}
It will not be too unfair to say that the proof of $\spadesuit$, in [DNR08], 
is not complicated. The possibility of having our favorite  instance $D$ fixed gives a lot
of control, and having this control it is not hard to encode a computation of a machine of one's choice as $\cal T$-chase for some program $\cal T$. 
The same can be said about $\diamondsuit$, whose proof, in [M09], 
is an adaptation of the proof of $\spadesuit$ -- the input instance over signature $\Sigma_1$ is neglected, 
a new instance, over $\Sigma_2$, hardwired in dedicated  TGDs, is created, and then the proof from [DNR08] is applied.\smallskip

\noindent
{\bf Flooding rule.} The schema from [DNR08] is repeated, in a sense, in the proof of $\heartsuit 2$ in  [GO13]. The instance $D$ is treated as 
a input of some machine, and chase simulates the computation of this machine on given input. Chase terminates when the computation does. The problem 
are the instances $D$ which contain too much positive information to be understood by a Datalog$^\exists$ program as a finite input -- for example
instances that contain a loop, which is unavoidably seen by a program as an infinite path.

The trick used in 
 [GO13] to  make sure that  chase will  terminate on such unwelcome instances
is the flooding rule -- a technique earlier used in 1990s in the numerous papers dealing with the Datalog boundedness problem [GMSV93]. Let us illustrate it by an example:\smallskip

\noindent
{\bf Example.} Consider the program $\cal T$:\smallskip

(i)~~~~ $U(x,y,z), E(z,w) \Rightarrow  \exists u\; U(y,u,w)$;

(ii)~~~ $E(x,y)\rightarrow E^+(x,y)$;

(iii)~~ $E^+(x,y), E(y,z) \Rightarrow E^+(x,z)$;

(iv)~~ $E^+(x,x) \Rightarrow U(y,u,w)$ \hfill (flooding rule)\smallskip

To see what is going on here, notice that $E$-atoms are never produced. Rules (ii) and (iii) compute $E^+$, being the (non-reflexive) transitive closure of $E$. 
Rule (i) unfolds graph $E$: if $\langle x,y\rangle $ is an edge 
in the unfolding,  $y$ is ``over''  an element $z$ in $E$ and if there is an edge $\langle z, w\rangle$ in $E$  then a new element $u$ 
must exist in  the unfolding, being ``over'' $w$. 

It is easy to see that whatever $D$ we begin with, $\cal T$-Standard Chase on $D$ has a terminating path. If $E$ is acyclic, then rule (i) terminates 
for all chase variants.
If $E$ has a cycle, then rules (ii) and (iii) can prove $E^+(a,a)$ for some $a$, and then rule (iv) can be used to ``flood'' the predicate $U$, so that 
in consequence, the head of (i) will be always satisfied and (i) will never be triggered again.

But there is no hope for this trick to work for the all--instances--all--paths Standard Chase: flooding rule only terminates a Standard Chase 
if we can make sure it is always used early enough to prevent new elements to be born, which means that it must be us who decides what the execution order is.

Clearly, this technique also fails for the eager chase variants.  $\cal T$-Oblivious Chase on $D$  does not terminate whenever $D$ is an instance containing an atom $U(a,b,c)$, for some $c$ belonging to a cycle in $E$. 

Notice also that adding a denial constraint to the constraints ($\clubsuit$) is just another way of using a flooding rule -- instead of flooding the database we make the chase fail.\smallskip

\noindent
{\bf Drinking from the well of positivity.} The trick we invented in this paper to replace the flooding rule is as follows. We treat the instance $D$ as
the only source of some positive facts: there are predicates which are never proved, they can only come with $D$. 

Then the idea is that each new element $a$ of Chase uses the path leading from $D$ to $a$ to  run its private computation 
of some Turing-complete  computational model. Only Datalog rules are used in this computation so we do not need to bother about termination. In order 
 to be able to
give birth to a successor  $a$ must first reach, by means of atoms  created during its private computation,  some atom that can only be found in $D$. 
Elements of Chase which are already too far away from this source of positivity cannot drink from it any more, dessicate, and do not
 produce offspring, thus causing the chase to terminate.

As we are going to see in the next Section,  once one knows the above
idea, the proof of Theorem \ref{glowne} is easy.

\section{Proof of Theorem 1}\label{dowod}

\subsection{The well of positivity}\label{well}
From now on, whenever we say ``chase'' we mean Oblivious Chase. 

Informally we say that Oblivious Chase creates one witness for each tuple satisfying the body of an existential TGD, 
regardless whether such a witness is already 
present in the current database instance or not. One of the ways how this informal statement can be formalized 
is to construct, for a given   Datalog$^\exists$ program $\cal T$  a new program ${\cal T}'$, by  replacing each 
$TGD$ in $\cal T$, of the form:\smallskip

(i) $\Phi(\bar x) \Rightarrow \exists y \; \Psi(y, \bar x)$\smallskip

where (i) is the number of the rule in $\cal T$, and $\Phi$ and $\Psi$ are conjunctive queries, by a rule:\smallskip

(i')  $\Phi(\bar x) \Rightarrow  \Psi(h_i(\bar x), \bar x)$\smallskip

where $h_i$ is a Skolem function. 
In this way  Chase($D, \cal T$) is the structure whose
active domain
 is a subset of Herbrand universe, where the elements of $D$ are treated as constants, and terms are built out of constants using the Skolem functions $h_i$, and 
which is a minimal model for all the rules of the program ${\cal T}'$. Since ${\cal T}'$ is a Prolog program it always has such a minimal model.

Now the question whether the $\cal T$-Oblivious Chase on $D$ terminates is equivalent to the question whether Chase($D, \cal T$),
seen as a 
substructure of the Herbrand universe, contains, for each $k\in \mathbb N$, a term of  depth at least $k$.

For a given signature $\Sigma$, an element  $a_\Sigma$ of a database instance $D$ over $\Sigma$ will be called {\em a well of positivity} if 
for each relation $R\in \Sigma$  the atom $R(a_\Sigma,a_\Sigma,\ldots a_\Sigma)$ is true in $D$. By $D_\Sigma$ we  will denote the database instance consisting of a single element,
being a well of positivity.

\begin{lemma}\label{studnia}
The following conditions are equivalent for  Datalog$^\exists$ program $\cal T$:\vspace{-3mm}

\begin{itemize}
\item[(i)] for each database instance $D$,   ~$\cal T$-Oblivious Chase on $D$ terminates;
\item[(ii)] $\cal T$- Oblivious Chase terminates on $D_\Sigma$.
\end{itemize}

\end{lemma}

This is (rephrased) Theorem 2 in [M09]. We sketch its proof for completeness.

\proof Only the (ii)$\Rightarrow$ (i) implication needs a proof. Let us
 assume that there exists $D$ such that  Chase($D, \cal T$),
seen as a 
substructure of the Herbrand universe, contains, for each $k\in \mathbb N$, a term of  depth at least $k$. What we 
need to prove is that also Chase($D_\Sigma, \cal T$) does contain such a term. 

So let $t$ be a term  of depth at least $k$ in Chase($D, \cal T$). This means that there is a derivation, in program ${\cal T}'$,
having atoms of $D$ in its leaves and some atom containing $t$ in its root. When we replace all the elements of $D$, occurring in atoms of this derivation,
by the well of positivity $a_\Sigma$, then we will get another valid derivation in program ${\cal T}'$, leading, instead of $t$, to some 
new term $t'$ in Chase($D_\Sigma, \cal T$). And the depth of $t'$ is equal to the depth of $t$ -- the two terms only differ at the level of constants,
but are equal otherwise. \eop

\subsection{The problem to be reduced}

The undecidable problem we are going to encode is the halting problem for finite automata with three counters (3CM). 
More precisely, the instance of the problem Halt3CM is a triple consisting of finite 
set $Q$ of states, of some initial state $q_1\in Q$ and of a finite set $\Pi$ of instructions, each of them of the following format:

\noindent
{\tt if the current state is $q\in Q$,\\
 the value of the first counter  (is$|$is not) zero\\
 and the value of the second counter (is$|$is not) zero\\
then:\\
change the state to $q'\in Q$;\\
(increment$|$decrement$|$keep unchanged) the value of the first counter,\\
(increment$|$decrement$|$keep unchanged) the value of the second counter,\\
increment the value of the third counter.}\medskip

We assume here that the automaton is deterministic, which means that the part of the instruction which is after {\tt then} is a 
 function of the part occurring before  {\tt then}. This function is partial -- if a configuration is reached with no instruction applicable then the automaton halts.

The problem, called Halt3CM, is  whether, for a given 3CM $M$,  executing the instructions of $M$  will ever halt when started from
the state $q_1$ and three empty counters. Of course  Halt3CM is undecidable. From now on each time  we say ``$M$ halts''  we   mean that 
it halts  after started from
 $q_1$ and three empty counters.

Notice that the value  of the third counter is never read by the automaton, and the counter is incremented in each step.
This leads to the following:

\begin{lemma}
A 3CM halts if and only if  the set of values of its third counter is bounded. 
\end{lemma}

From now on a 3CM $M=\langle Q,q_1,\Pi\rangle$ is fixed and we will construct a Datalog$^\exists$ program ${\cal T}_M$, over some signature 
$\Sigma_M$  such that 
   ${\cal T}_M$-Oblivious Chase   on $D_{\Sigma_M}$ terminates if and only if $M$ halts.

\subsection{Encoding the automaton as a Conway function}

Now we will encode the computation of $M$  as a sequence of iterations of a Conway function. This technique is by no means new, 
but maybe not as widely known as some other undecidable problems, so we include this subsection for  completeness.

Suppose $|Q|=m$. Let $p_1=2$, $p_2=3,\ldots$ $p_{m+3}$ be the first $m+3$ primes and let $p=p_1p_2\ldots p_{m+3}$.
Let $\bf c$ be 
 a configuration of $M$ with the state being $q_i$ and $c_1$, $c_2$ and $c_3$ being respectively values of the first, second and third counter.
Then by $e({\bf c})$ (or {\em encoding of}  $\bf c$) we will mean the number:

$$p_ip_{m+1}^{c_1}p_{m+2}^{c_2}p_{m+3}^{c_3}$$

Notice that if  ${\bf c}$ is the initial configuration of $M$ then   $e({\bf c})=2$.

For two configurations  $\bf c$,  ${\bf c}'$ of $M$ we will say that they are consecutive when ${\bf c}'$ is a result of executing a single step 
of $M$ in  ${\bf c}$ or when there is no instruction that can be executed in  ${\bf c}$ and  ${\bf c}={\bf c}'$.
Now it is easy to see that:

\begin{theoremm}\label{conway}
There exist natural numbers $q_0$, $q_1$,\ldots $q_{p-1}$,  $r_0$, $r_1$,\ldots $r_{p-1}$, such that for each two consecutive configurations 
$\bf c$,  ${\bf c}'$ of $M$, such that $e({\bf c}) = i\mod  p $ it holds that $e({\bf c}')= \frac{q_ie({\bf c})}{r_i} $. 
\end{theoremm}

For the proof of this theorem notice that the reminder $i$ of  $e({\bf c})$ modulo $p$ carries all the information needed for $M$ to decide 
which instruction should be applied: the state is $q_j$ if and only if $i$ is divisible by $p_j$ and the value of the (for example) 
second counter is non-zero if
and only if $i$ is divisible by $p_{m+2}$. It is equally easy to see that executing an instruction boils down to division (removing the old state, 
decrementing a counter) and multiplication (moving to a new state, incrementing a counter).

From now on the numbers  $q_0$, $q_1$,\ldots $q_{p-1}$,  $r_0$, $r_1$,\ldots $r_{p-1}$ provided for $M$ by  Theorem \ref{conway} are fixed. Denote by
$g$ a function that maps a natural number $n$ to $nq_i/r_i$, where   $ n=i\mod  p $. Let ${\cal G}=\{g^n(2): n\in\mathbb N\}$ be the smallest subset of $\mathbb N$ 
which contains 2 and is closed under $g$. Clearly,
$M$ halts if and only if $\cal G$ is bounded. So, what remains for us to do is to  construct such a Datalog$^\exists$ program ${\cal T}_M$ that 
 ${\cal T}_M$-Oblivious Chase on $D_{\Sigma_M}$ terminates if and only if $\cal G$ is bounded. Notice that it is here where the third counter is important.

\subsection{The program ${\cal T}_M$}

Denote by $\cal QR$ the set $\{q_0, q_1,\ldots q_{p-1}, r_0, r_,\ldots r_{p-1}\}$.
The signature $\Sigma_M$ will consist of the following relations:

\begin{itemize}
\item a binary relation $E$, which will pretend to be the successor relation on the natural numbers;

\item for each $j\in \cal QR$ a binary relation 
$E^j$ -- only needed to keep rule (d3) short;

\item a unary relation $H$, which will never occur in the head of any rule, so its only atom will  be $H(a_\Sigma)$;

\item for each $0\leq i\leq p-1$ a ternary relation $T^i$, 
with $T^i_x(y,z)$ meaning something like ``$x$ thinks that $\frac{y}{z} = \frac{q_i}{r_i}$''. Normally we should of course  write 
$T(x,y,z)$ rather than $T_x(y,z)$. But we like $T_x(y,z)$ more, and it is still ternary;

\item for each $0\leq i\leq p-1$ a binary relation $R^i$, with $R^i_x(y)$ meaning something like ``$x$ thinks that $i = y \mod p $'';

\item a binary relation $G$, with $G_x(y)$ meaning ``$x$ thinks that $y\in \cal G$'';

\item a unary relation $N$, with $N(x)$ meaning that  $x$ is a natural number. $N$ is not really needed, we only have it because otherwise the bodies
of rules (d2) and (d4) would be empty, and 
we do not like rules with empty bodies.
 
\end{itemize}

Now we are ready to write the program ${\cal T}_M$. 
There is one existential rule:\smallskip

\noindent
(e) $ G_x(y), H(y) \Rightarrow \exists  z \; E(z,x)$.\smallskip

\noindent
Read this rule  as ``Once $x$ has drunk from the well of positivity, it is allowed to give birth to a new element $z$.''\smallskip

There will be also several Datalog rules: \smallskip

\noindent
(d0) $E(y,y_1), E(y_1,y_2),\ldots E(y_{j-1},y_{j})\Rightarrow E^j(y,y_j)$  \hfill one rule for each  $j\in \cal QR$;\smallskip

\noindent
(d1) $E(z,x)  \Rightarrow N(z) $\medskip

\noindent
Rules of the form (d2) and (d3) form a recursive definition of multiplication 
by addition (remember -- $x$ always thinks it equals zero):\smallskip

\noindent
(d2) $N(x) \Rightarrow T^i_x(x,x)$ \hfill one rule for each  $0\leq i\leq p-1$;\smallskip

\noindent
(d3)
 $T^i_x(y,z), E^{q_i}(y,y'), E^{r_i}(z,z')  \Rightarrow T^i_x(y',z')$ \hfill one rule for  each $0\leq i\leq p-1$;\smallskip

\noindent
The next two rules count modulo $p$:\smallskip

\noindent
(d4) $N(x) \Rightarrow R^0_x(x)$;\smallskip

\noindent
(d5) $R^i_x(y), E(y,y') \Rightarrow R^j_x(y')$ \hfill  whenever $j=i+1 \mod p$;\smallskip

Now, once we have all the  predicates we need for the multiplications, and for remainders modulo $p$, we can easily write rules which will compute the 
set $\cal G$. First of them says -- as long as $x$ keeps assuming that it equals zero -- that $2\in \cal G$:\smallskip

\noindent
(d6) $E(x,y), E(y,z)\Rightarrow G_x(z)$\smallskip

Second rule for $G$  says that $\cal G$ is closed with respect to the function $g$:\smallskip

\noindent
(d7) $R^i_x(y), G_x(y), T^i_x(y,z) \Rightarrow G_x(z)$   \hfill one rule for  each $0\leq i\leq p-1$;\smallskip

Notice that the rules (d2)--(d7) form a sort of a private Datalog program for each $x$, 
and the atoms proved by such programs for different $x$, $x'$ never see each other
(this is reflected in our notation, which suggests that $x$ is more than merely an argument of the predicates, but part of their names). 
Rule (e) creates a new element $z$, such that  $E(z,x)$,  when the program for $x$ can prove that $ G_x(y)$ for some  $y$ such that $H(y)$. But, as we said,
 there is no rule saying that something is in $H$ and the only element $a$ such that 
Chase$(D_{\Sigma_M},{\cal T}_M) \models H(a)$ is the well of positivity $a_{\Sigma_M}$. So (e) creates a new element $z$, such that  $E(z,x)$,  when the program for $x$ can prove that 
$ G_x(a_{\Sigma_M})$. \smallskip

Now we have a lemma that  Theorem \ref{glowne} follows from:

\begin{lemma}\label{rownowaznosc}
  ${\cal T}_M$-Oblivious Chase on $D_{\Sigma_M}$ terminates if and only if $\cal G$ is bounded.
\end{lemma}

We think that the lemma follows directly from the construction of  ${\cal T}_M$. But the readers who like it more formal, are invited to read Appendix A.

\subsection{The case of all-instances-all-paths Standard Chase termination}\label{standard}

For any $\cal T$ and $D$ any structure being a result of running a $\cal T$-Standard Chase on $D$ is a subset of (oblivious) Chase$(D, {\cal T})$.
This means that if $\cal G$ is bounded, then   ${\cal T}_M$-Standard Chase terminates on each instance and each path. What remains to be seen is that if
 $\cal G$ is not bounded, then there exists $D$ such that  ${\cal T}_M$-Standard Chase does not terminate on some path. It is easy to see that
a structure $D$, consisting of the well of positivity $a_{\Sigma_M}$ and of some $a$ such that $D\models E(a,a_{\Sigma_M})$, has  this property.

\section{Proof of Theorem \ref{glowne3}}\label{bibarne}
\vspace{-1.5mm}

It is  harder to prove any nontrivial lower bound for all-instances Oblivious Chase  termination problem for single-head TGDs over binary signatures, 
then to prove undecidability in the general case. In the proof of Theorem \ref{glowne3} we try to repeat the  idea of proof of Theorem \ref{glowne}, 
creating a new element of some $E$-path, for a binary  $E$, only when some private computation,
 run by the last element $a$ of the current path,  terminates. But, while having  arity three at our disposal, we could run many mutually non-interfering computations 
using the same arena, now we must construct a separate arena for each element of the $E$-path being built. 

This arena needs to be huge enough to contain a complex computation, but on the other hand the process of the construction of the arena should never lead to 
an infinite chase. In other words we need to  -- and we think it is not 
 immediately clear 
how to do it -- find  a binary Datalog$^\exists$ program which builds a huge 
(i.e. greater than exponential, with respect to  the size of the program) 
(Oblivious) Chase, when run on $a_\Sigma$, 
but  finally terminates.   \vspace{-1mm}

\subsection{Constructing the arena: Chase of exponential depth} \vspace{-1mm}

Let $m$ be a fixed natural number and let $M=2^m$.  
Consider the program ${\cal T}^0_b(m)$ consisting of the following rules:\\

(d0) $H(x)\Rightarrow K(x)$\smallskip

(d0') $H(x)\Rightarrow C_i(x)$ \hfill (one rule for each $i\in\{0, 1, \ldots m\}$)\smallskip

(e) ~$K(x) \Rightarrow \exists y \; R(x,y)$\smallskip

(d1) $R(x,y) \Rightarrow  T(y,y)$\smallskip

(d2) $T(x,y), R(x',z), R(z, x), R(y',y)\Rightarrow T(x',y')$\smallskip

(d3) $T(x,y), C_i(x) \Rightarrow C_{i+1}(y)$ \hfill (one rule for each $i\in\{0, 1, \ldots m-1\}$)\smallskip

(d4) $R(x,y),  C_m(x)\Rightarrow K(y)$\smallskip

Let now $a$ be any element such that $H(a)$ (which means that $a$ may be, but may not be,  a well of positivity),
and let $D_a$ be a database instance containing $a$ as a single element. 

\begin{exercisee}\label{cwiczenie}
Chase$(D_a,{\cal T}^0_b(m))$, seen as a graph over predicate $R$, is a path of length $M+1$, having $a$ as its  first element
\end{exercisee}

Solution to this exercise can be found in Appendix B. {\bf Hint:} like in Section \ref{dowod}  there is no rule saying that something is in $H$, and the
only element satisfying $H$ plays the role of the well of positivity. Also like in Section \ref{dowod}, Oblivious Chase produces a path (this time it is an $R$-path) -- 
if an element is in $K$ then it is ``close enough'' to $H$ to be able to produce
$R$-offspring. The predicates $C_i$ are resources -- the further we are from $H$ the more we are running out them.\vspace{-1mm}

\subsection{Constructing the arena: Chase of double exponential size} \vspace{-1mm}

For fixed natural numbers $m$ and  $p$ 
consider now the program ${\cal T}^1_b(m,p)$ consisting of all the rules that can be obtained from the rules 
of  ${\cal T}^0_b(m)$ by replacing each occurrence of the predicate $R$ with one of the predicates $R_1, \ldots R_p$.
For example
rule (e) will be replaced by $p$ new rules while
rule (d2) will be replaced by $p^3$ new rules. Let  $a$  and  $D_{a}$ be as in the previous subsection.
Then the analysis of Chase$(D_{a},{\cal T}^1_b(m,p))$ is analogous to the analysis of Chase$(D_a,{\cal T}^0_b(m))$, except that the structure we now get  is a  
 $p$-ary tree of depth $M+1$ rather than  a path of length $M+1$.  Notice that the same elements are created  regardless if $a$ is a well of positivity, or any
element just satisfying $H(a)$.

\subsection{The encoding Lemma and how it implies Theorem \ref{glowne3}} \vspace{-1mm}

Now Chase$(D_{a},{\cal T}^1_b(m,p))$  can be used as an arena, where we can run some computation. Let $a$ and $D_a$ be as before. \vspace{-1mm}

\begin{lemma}[The encoding Lemma]\label{encodinglemma} The problem:

\shaded{
Given $m,p\in \mathbb N$ and a Datalog program $\cal T$, with EDB relations $H$, $R_1$, $R_2,\ldots R_p$
 and IDB relations 
$P$ (binary) and  $G_1$,$G$, $G_2$  and $C$ (unary). Is it the case that:
 $$ Chase(D_a, {\cal T}^1_b(m,p)\cup {\cal T}) \models C(a)\;\;\;?$$}

\noindent
is EXSPACE-hard.\\ The size of the instance is here the size of the program ${\cal T}^1_b(m,p)\cup {\cal T}$.
\end{lemma}

\noindent
For the proof of the Lemma see Appendix C. 
 Notice that  $Chase(D_a, {\cal T}^1_b(m,p))\cup {\cal T})$ has the same set of elements as 
Chase$(D_a,{\cal T}^1_b(m))$ -- this is because the  Datalog rules of $\cal T$ do not prove any atoms that could be used by ${\cal T}^1_b(m,p)$.\smallskip

\noindent
Let now 
${\cal T}^2_b(m,p)$ be ${\cal T}^1_b(m,p)$ with the following additional rules:\smallskip

(d') $ E(x,y)\Rightarrow H(y)$

(e') $C(x) \Rightarrow \exists z \; E(x,z)$.\smallskip

\noindent
Proof of Theorem \ref{glowne3} will be finished when we show:

\begin{lemma}

For a Datalog program $\cal T$, as in Lemma \ref{encodinglemma},  the following two conditions are equivalent:\vspace{-1mm}
\begin{itemize}
\item $ Chase(D_a, {\cal T}^1_b(m,p)\cup {\cal T}) \models C(a)$

\item $ Chase(D_\Sigma, {\cal T}^2_b(m,p)\cup {\cal T}$ does not terminate.
\end{itemize}
\end{lemma}

For the proof of the Lemma first suppose that  $ Chase(D_a, {\cal T}^1_b(m,p)\cup {\cal T}) \models C(a)$. 
Let us run $ {\cal T}^2_b(m,p)$ on $D_\Sigma$.
Since $C(a_\Sigma)$ is true in $D_\Sigma$,
rules (e'), and then (d') will be triggered, creating a new element $c$ satisfying $H(c)$. 
Then, rules of  ${\cal T}^1_b(m,p)$ will build  the p-ary tree of depth $M+1$ rooted in $c$ and ${\cal T}$ 
will be run on this tree, proving 
$C(c)$. But this means that  (e') will trigger again, creating element $c'$ such that $E(c,c')$ and $H(c')$, and so on.

Now suppose that  $ Chase(D_a, {\cal T}^1_b(m,p)\cup {\cal T}) \not\models C(a)$. Then again, an element $c$ like above will be created,
and the p-ary tree of depth $M+1$ rooted in $c$ will be built, $\cal T$ will be run on this tree, but $C(c)$ will never be proved,  no new elements will be added,
and chase will terminate.\eop \vspace{-1mm}

\section{References}  \vspace{-1mm}

{\small
\noindent
[ASU79] A.V. Aho, Y. Sagiv, J.D. Ullman,  {\em Efficient Optimization of a Class of Relational Expressions};
ACM Transactions on Database Systems, 4(4):435--454, 1979;

\noindent
[B84] C. Beeri, M. Y. Vardi. {\em A proof procedure for data dependencies}; Journal of the ACM (JACM) 31.4 (1984): 718-741.

\noindent
[CGK08] A. Calı, G. Gottlob, M. Kifer, {\em Taming the infinite chase: Query answering under expressive relational constraints};
 Proc. of KR (2008): 70-80.

\noindent
[CGL09] A. Calì, G. Gottlob, T. Lukasiewicz, {\em "Datalog +/-: a unified approach to ontologies and integrity constraints}; Proceedings of the 12th International Conference on Database Theory. ACM, 2009.

\noindent
[CGL12] A. Calì, G. Gottlob, T. Lukasiewicz, {A general datalog-based framework for tractable query answering over ontologies}; Web Semantics: Science, Services and Agents on the World Wide Web 14 (2012): 57-83.

\noindent
[DT03] A. Deutsch, Val Tannen, {\em Reformulation of XML queries and constraints}; Database Theory—ICDT 2003. Springer Berlin Heidelberg, 2002. 225-241.

\noindent
[DNR08] A. Deutsch, A. Nash,  J. Remmel,  {\em The chase revisited}; Proc. of the 27th ACM SIGMOD-SIGACT-SIGART symposium on Principles of database systems. ACM, 2008.

\noindent
[FKPP05] R. Fagin, P.G. Kolaitis,  R. J. Miller, L. Popa, {\em Data exchange: semantics and query answering}; Theoretical Computer Science, 336(1), 89-124 (2005).

\noindent
[G10] S. Greco,  F. Spezzano, {\em Chase termination: A constraints rewriting approach}; Proceedings of the VLDB Endowment 3.1-2 (2010): 93-104.

\noindent
[GMSV93]
H. Gaifman 	H. Mairson 	
	Y. Sagiv 	
	M. Y. Vardi {\em Undecidable optimization problems for database logic programs} Journal of the ACM 
Volume 40:3; 1993, 683-713.

\noindent
[GO11] G. Grahne,  A. Onet. {\em On Conditional Chase Termination}; AMW 11 (2011): 46.

\noindent
[G013] G. Grahne, A. Onet {\em Anatomy of the chase};   arXiv:1303.6682 (2013).

\noindent
[HS07] A. Hernich,  N. Schweikardt {\em CWA-solutions for data exchange settings with target dependencies};
Proceedings of the twenty-sixth ACM SIGMOD-SIGACT-SIGART symposium on Principles of Database Systems. ACM, 2007.
%

\noindent [M99] J. Marcinkowski,
{\em
Achilles, turtle, and undecidable boundedness problems for small DATALOG programs};
SIAM Journal on Computing
29:1, 1999,
231-257.

\noindent
[MMS79] D. Maier,  A. O. Mendelzon, Y. Sagiv. {\em Testing implications of data dependencies}; ACM Transactions on Database Systems  4.4 (1979): 455-469.

\noindent
[M09] B. Marnette,  {\em Generalized schema-mappings: from termination to tractability}; Proceedings of the twenty-eighth ACM SIGMOD-SIGACT-SIGART symposium on Principles of Database Systems. ACM, 2009.

\noindent
[MSL09] M. Meier, M. Schmidt, and G. Lausen, {\em On chase termination beyond stratification}; Proceedings of the VLDB Endowment 2.1 (2009): 970-981.

\noindent
[MSL09a] M. Meier,  M. Schmidt,  G. Lausen,  {\em On chase termination beyond stratification} [technical report and erratum] Website. 
{\tt http://arxiv.org/abs/0906.4228}

\noindent
[O13] A. Onet, {The chase procedure and its applications in data exchange};  Data Exchange, Integration, and Streams. Dagstuhl Follow-Ups, Schloss Dagstuhl-Leibniz-Zentrum für Informatik, Germany, 2013.

\newpage

\section{Appendix A. Proof of Lemma \ref{rownowaznosc}}

As there is no risk of confusion we will denote the structure, being the result of  ${\cal T}_M$-Oblivious Chase on $D_{\Sigma_M}$   simply as Chase.

We are going to prove two lemmas describing the  structure of Chase. 
First, let us think of Chase as of a graph 
with respect to the relation $E$. Since we do not want to be distracted by the loop $E(a_{\Sigma_M},a_{\Sigma_M})$ all the time, let $E_0$ be the relation $E$ in Chase
minus the edge $E(a_{\Sigma_M},a_{\Sigma_M})$.

\begin{lemma}\label{sciezka}
Chase is a descending $E_0$-path, finite or infinite, without self-loops. 
The first element of this path  is  the well of positivity $a_{\Sigma_M}$. 
\end{lemma} 

In order to prove Lemma \ref{sciezka} it is enough to show that:

\begin{enumerate}
\item[(i)] $E_0$ is a connected graph;
\item[(ii)] there are no $E_0$-cycles in Chase,
\item[(iii)] the $E_0$-out-degree of each element of Chase is 1, except from  $a_{\Sigma_M}$, whose out-degree is 0;
\item[(iv)] the $E_0$-in-degree of each element of Chase is at most 1.
\end{enumerate}

To see that (i) holds true notice that for each element $a$ of Chase, there exists a (descending) $E_0$ path from  $a_{\Sigma_M}$ to $a$. 
It can be easily proved by induction on the structure of Chase that this path already exists at the moment when $a$ is created.

For the proof of (ii) notice that whenever Chase $\models E_0(a,b)$ then $b$ was created by the chase procedure earlier than $a$. 

Concerning (iii), notice that the only way for an element $b\neq a_{\Sigma_M}$ to be in an atom $E_0(a,b)$ in Chase, for some $a$, is
to be created by rule (e) from elements $x=a$, $y=a_{\Sigma_M}$.  

For claim (iv) it is enough to see that the only rule of ${\cal T}_M$ that creates atoms of $E$  is rule (e). Since there is only one element satisfying 
$H$, each element of Chase is involved, as the variable $x$, in at most one tuple satisfying the body of rule (e). By the rules of oblivious chase 
this means that rule (e) is triggered at most once for each element of Chase being $x$. This ends the proof of Lemma \ref{sciezka}. \eop

Let now $a=a_0$ be any element of Chase and for each $i\in \mathbb N$
let $a_{i+1}$ be the unique element of Chase such that Chase $\models E(a_i,a_{i+1})$. Of course there exists $k\in \mathbb N$ such that 
for each $i\geq k$ we have $a_i = a_{\Sigma_M}$. Let $k_0$ be the smallest such $k$. 
Then the following lemma follows easily from Lemma \ref{sciezka}  and from the construction 
of rules (d2)--(d7): 

\begin{lemma}\label{trywialny}
\begin{enumerate}
\item[(i)] If there is an $E$-path in Chase from $a$ to some $b$ then $b=a_j$ for some $j\in \mathbb N$;

\item[(ii)] if $T^i_a(b,c)$ holds in Chase for some elements $b,c$ then $b=a_j$, $c=a_{j'}$ for some $j,j'\in \mathbb N$ such that $jr_i=j'q_i$;

\item[(iii)] if $R^i_a(b)$ holds in Chase for some element $b,c$ then  $b=a_j$ for some $j\in \mathbb N$ such that $j=i \mod p$;

\item[(iv)] if $\cal G$ is bounded, and $k_0$ is greater than all the elements of $\cal G$ then ${\cal G} = \{j: $ Chase $\models G_{a}(a_j)\}$

\item[(v)] if $\cal G$ is unbounded then ${\cal G} \subseteq \{j: $ Chase $\models G_{a}(a_j)\}$; in particular in such case Chase $\models G_{a}(a_{\Sigma_M})$.

\end{enumerate}
\end{lemma}

Notice that if $\cal G$ is unbounded then it may be that Chase $\models G_{a}(a_j)$ for some $j\not\in \cal G$, even if $j<k_0$. 
We imagine the predicate $G_a$ as $a$ moving a pebble to the values of subsequent iterations of $g$. But once the pebble falls to the well of positivity
all the control is lost, and different things can happen. 

Now, we are ready to prove Lemma \ref{rownowaznosc}. 

Suppose first that  $\cal G$ is unbounded. We will show that in such case Chase is an infinite descending $E$-path, which means that for each $a$ in Chase there is a $b\neq a$ such that Chase $\models E(b,a)$. But by Lemma \ref{trywialny} (v)
if $\cal G$ is unbounded then Chase $\models G_{a}(a_{\Sigma_M})$, so the body of rule (e) is satisfied in Chase 
for $x=a$ and its head must also be satisfied.

Now suppose that   $\cal G$ is bounded, that $k$ is a natural number greater than all the elements of $\cal G$, 
 and that Chase is an infinite path. Let $a$ be any element such that the $E$-distance between $a$ and
$a_{\Sigma_M}$ is greater than $k$. Then, by  Lemma \ref{trywialny} (iv) $G_{a}(a_{\Sigma_M})$ is never proved, 
and 
the rule (e) could never have been triggered for $a_{0}$ as $x$. But this contradicts the assumption that Chase was infinite.\eop

\section{Appendix B. Solution to Exercise \ref{cwiczenie} }

It is clear that Chase$(D_a,{\cal T}^0_b(m))$, seen as an $R$-graph is a path: this is because 
rule (e) can only create one $R$-successor for each node.

Now, suppose that in the process of building Chase$(D_a,{\cal T}^0_b(m))$   we always trigger Datalog
rules as early as possible, and that rule (e) is only used when there are no more Datalog rules applicable 
(this can be assumed since -- as we already 
noticed in the Introduction -- 
the order in which rules are used by Oblivious Chase does not matter). Let  ${\cal C}_N$ be the 
(partial) Chase$(D_a,{\cal T}^0_b(m))$ after rule (e) was used for the $N$th time, which means that
 ${\cal C}_N$ consists of $N+1$ elements, call them
$a_0=a$, $a_1$, $a_2, \ldots a_N$ (in the order that they were created in), and after all
the Datalog rules were saturated. Notice that we do not claim that ${\cal C}_N$ exists for each $N$.

\begin{lemma}\label{ote}
${\cal C}_N \models T(a_i,a_j)$ if and only if $i+N\leq 2j$ 
\end{lemma}

\proof Easy induction -- on the depth of the derivation.\eop

\begin{lemma}\label{oce}
Suppose $1\leq i\leq m$. Then ${\cal C}_N \models C_i(a_j)$ if and only if 
$\frac{j}{N}\leq \frac{2^i-1}{2^i}$.
\end{lemma}

\proof Induction on $i$. Use Lemma \ref{ote}.\eop

\begin{lemma}\label{oka}
${\cal C}_N \models K(a_j)$ if and only if $\frac{j-1}{N}\leq \frac{2^m-1}{2^m}$.
\end{lemma}

\proof
 Notice that
(unless $j=0$) the only rule that can prove  $K(a_j)$ is rule (d4). This means that 
${\cal C}_N \models K(a_j)$ if and only if ${\cal C}_N \models C_m(a_{j-1})$. But -- due to Lemma   \ref{oce}
this holds if and only if $\frac{j-1}{N}\leq \frac{2^m-1}{2^m}$.\eop

\begin{lemma}\label{oen}
${\cal C}_N \models K(a_N)$ if and only if $N\leq M$
\end{lemma}

\proof Lemma \ref{oka} says that 
${\cal C}_N \models K(a_N)$ if and only if $\frac{N-1}{N}\leq \frac{2^m-1}{2^m}$.\eop

But this means that rule (e) will be triggered  in ${\cal C}_N$ if and only if $N\leq M$. Which ends the
solution of Exercise \ref{cwiczenie}.\eop

\section{Appendix C. Proof of Lemma \ref{encodinglemma} (the encoding Lemma)} 

An instance of the problem we are going to encode, call it Thue$^2_2$, consists of:

\begin{itemize}
\item a finite  ${\cal A}= \{1,2,\ldots p\}$;
\item a natural number $m$;
\item a set of productions $\pi \subseteq {\cal A}^2 \times {\cal A}^2$.
\end{itemize}

Notice that each production of our Thue process replaces an infix of length 2 by another infix of length 2. So obviously,
the word problem is decidable for Thue$^2_2$. It is however straightforward to prove, by a standard encoding of a Turing machine, 
that the problem:

\shaded
{Given an instance of  Thue$^2_2$. Does this instance have a solution, which means that 
there exists a number $k< M$ such that $1p^k \stackrel{\ast}{\longrightarrow}_{\pi} 2p^k$~~?}

is EXPSPACE-complete. Notice that, as always, $M=2^m$.\smallskip

Let us remind the reader that $1p^k \stackrel{\ast}{\longrightarrow}_{\pi} 2p^k$ means that, beginning from some 
word of the form $1p^k$,
one can reach the word  $2p^k$ in some number of steps, in each step replacing some infix $w$ of a current word by an infix $w'$ in such a way that 
$\langle w, w'\rangle\in \pi$. Notice that if $1p^k \stackrel{\ast}{\longrightarrow}_{\pi} 2p^k$ is true for some $k<M$ then also 
$1p^{M-1} \stackrel{\ast}{\longrightarrow}_{\pi} 2p^{M-1}$ is true, so the  statement of the problem may seem to be unnecessarily   complicated. 
 But this is how we need it. 

From now on we assume that an instance $\Pi=\langle p,m,\pi\rangle  $ of Thue$^2_2$ is fixed. 
Our goal is to build a 
 Datalog program $\cal T$, over the signature as required by Lemma \ref{encodinglemma}
and 
such that 
$ Chase(D_a, {\cal T}^1_b(m,p))\cup {\cal T}) \models C(a)$ 
 if and only if $\Pi$ has a solution.

Let us remind the reader that  
Chase$(D_a, {\cal T}^1_b(m,p))$ 
is a tree, and elements of this tree can be in a natural way seen as words from  ${\cal A}^{\leq M}$.
The program $\cal T$ will first of all contain the the following rules, defining some new binary relation $P$ on Chase$(D_a, {\cal T}^1_b(m,p))$
 (seen as ${\cal A}^{\leq M}$):\\

\noindent
(p1) $ R_i(x,y), R_{i'}(y,y'), R_j(x,z), R_{j'}(z,z')\Rightarrow P(y',z') $  \hfill   one rule for each\\ 
\hspace*{94mm}  pair $\langle ii',jj'\rangle \in \pi$.\medskip

\noindent
(p2) $P(x,y), R_i(x,x'), R_i(y,y') \Rightarrow P(x',y')$\hfill   one rule for each $i\in \cal A$.\medskip

It is easy to see that the predicate $P$  computes pairs of words $\langle w,w'\rangle \in {\cal A}^{\leq M}\times {\cal A}^{\leq M}$ such that 
$w {\longrightarrow}_{\pi} w'$, which means that $w$ rewrites to $w'$ by the Thue process $\Pi$ in exactly one step. Rule (p1) says 
that whenever there are two $a$ and $b$ elements of Chase$(D_a, {\cal T}^1_b(m,p))$, which represent
 words of the form $wii'$ and $wjj'$, such that  $\langle ii',jj'\rangle \in \pi$ then $P(a,b)$ holds true in 
$Chase(D_a, {\cal T}^1_b(m,p))\cup {\cal T})$. 
 Rule 
(p2) provides a mechanism able to add any (but the same) suffix, both to $wii'$ and $wjj'$.

Now we are going to play a pebble game, like in Appendix A.
Next  three rules of $\cal T$  allow us to place a pebble on any element of Chase$(D_a, {\cal T}^1_b(m,p))$   that represents 
a word of the form $1p^k$ for some $k\leq M$:\smallskip

\noindent
(g1) $H(x), E_1(x,y)\Rightarrow G^1(y)$;\smallskip

\noindent
(g2) $G^1(y), E_p(y,y') \Rightarrow  G^1(y')$;\smallskip

\noindent
(g3) $ G_1(x) \Rightarrow G(x)$.\smallskip

The next rule is the main mechanism of $\cal T$. It lets us follow the derivation in $\Pi$:\smallskip

\noindent
(g4) $G(y), P(y,y')\Rightarrow G(y')$.\smallskip

Finally, there are three rules in $\cal T$ that make it possible  check whether 
the element that we placed the pebble on does indeed represent a word of the form  $2p^k$:\smallskip

\noindent
(g5) $G(y) \Rightarrow G_2(y) $;\smallskip

\noindent
(g6) $G_2(y), E_p(y',y) \Rightarrow G_2(y') $\smallskip

\noindent
(g7) $H(x), E_2(x,y), G_2(y)\Rightarrow C(x)$.\smallskip

It now follows from the construction that 
Chase$(D_{a},{\cal T}^1_b(m,p)\cup {\cal T} \models C(a)$ if and only if $\Pi$ has a solution.

\section{\hspace{-2.9mm}Appendix D. Low arity vs. single head TGDs. Discussion.}

In the program ${\cal T}_M$ we constructed in Section \ref{dowod} there are several ternary rules of the form:\\

\noindent
(*) $T(x,y,z), E^q(y,y'), E^r(z,z')  \Rightarrow T(x,y',z')$\\

One could think that, if we allowed multi-head TGDs, the ternary relation in the head of rule (*) could be easily split into three binary relations:\\

\noindent
(**)    $T_1(v,x), T_2(v,y), T_3(v,z), E^q(y,y'), E^r(z,z') \Rightarrow\exists w\;  T_1(w,x), T_2(w,y'), T_3(w,z')$\\

where $v$ and $w$ are ``names'' for atoms $T(x,y,z)$ and $ T(x,y',z')$.
Of course all other rules involving $T$  would also need to be changed accordingly.\\

This is however not so simple. Notice how careful we were, in Section \ref{dowod}, about existential TGDs. There was only one of them, and not easy to trigger.
And, while  replacing (*) with (**), we replace a safe Datalog rule with a potentially prolific existential one. Indeed, it is not hard to see that, if we consider
the Oblivious Chase,  the $w$ in the head of (**) depends, as a Skolem term, on the $v$ in the body, which results with recursive calls and 
infinite chase. However, in the case of the 
Semi-Oblivious Chase, $v$ only depends on $x$, $y'$ and $z'$, and the only candidates for $y'$ and $z'$ are  all the elements on the 
$E$-path from the current $x$ to  $a_\Sigma$, which are finitely many. This means that using the above splitting we can really rewrite proof from Section \ref{dowod}
to a proof of:

\begin{theoremm}
All-instance termination of Semi-Oblivious Chase is undecidable (and r.e.-hard) 
 for programs consisting of multi-head TGDs over binary signatures.
\end{theoremm}

  As we said in the Introduction, we do not know
whether all--instances--all--paths Standard Chase termination is recursively enumerable.  Actually, it could very well be the case 
that it is r.e. for single-head TGDs but not for multi-head TGDs. 
In particular,  the standard translation of multi-head TGDs into single head TGDs, where a new predicate is added for the head of each rule, from which
the atoms of the head are then produced using projections, does not preserve all-instances-all-paths Standard Chase termination.  As an example
consider a program $\cal T$ consisting of a single rule:\smallskip

~~~~$E(x,y)\Rightarrow \exists z\; E(y,z), E(z,y)$\smallskip

$\cal T$-Standard Chase on $D$ terminates for each $D$. Let however the following ${\cal T'}$ be the natural translation of ${\cal T}$:\medskip

~~~~$E(x,y)\Rightarrow \exists z\; R(y,z)$ \smallskip

~~~~$R(y,z)\Rightarrow E(y,z) $ \smallskip

~~~~$R(y,z)\Rightarrow  E(z,y)$ \medskip

Then it is easy to see that   ${\cal T}'$-Standard Chase  does not always terminate.

\end{document}